\documentclass[review,5p]{elsarticle}
\usepackage[T1]{fontenc}
\usepackage[utf8]{inputenc}
\usepackage{lmodern, graphicx, prettyref, amsmath, physics}
\usepackage{siunitx}
\usepackage[version=4]{mhchem} 
\usepackage{float}
\usepackage{hyperref}
\usepackage[nameinlink, capitalize]{cleveref}

\makeatletter
\def\ps@pprintTitle{%
 \let\@oddhead\@empty
 \let\@evenhead\@empty
 \def\@oddfoot{}%
 \let\@evenfoot\@oddfoot}
\makeatother

\DeclareSIUnit{\rydberg}{Ry} \DeclareSIUnit{\hopernmsquared}{HO\per\nm\squared}
\DeclareSIUnit{\kt}{k_BT}

\graphicspath{{figures/}}

\begin{document}

\begin{frontmatter}

\title{\emph{Ab initio} molecular dynamics description of proton transfer at water-tricalcium silicate interface}

\author[add1,add2]{Jérôme Claverie}
\author[add1]{Fabrice Bernard}
\author[add2]{João Manuel Marques Cordeiro}
\author[add1]{Siham Kamali-Bernard}
\ead{siham.kamali-bernard@insa-rennes.fr}
\address[add1]{Laboratory of Civil Engineering and Mechanical Engineering (LGCGM), Rennes University, INSA Rennes, Rennes, France} \address[add2]{Department of Physics and Chemistry, School of Natural Sciences and
				Engineering, São Paulo State University (UNESP), 15385-000 Ilha
				Solteira, São Paulo, Brazil}

\begin{abstract}
For the first time, an \emph{ab initio} molecular dynamics simulation was
performed to describe the \ce{C3S}/water interface. The simulation shows that
oxides with favorable environment are protonated at first, creating very stable
hydroxide groups. Proton transfers occur between water and silicates, and
between water and hydroxides formed upon water dissociation on the surface.
The typical lifetime of these events is on the same timescale than interconversion between Eigen and Zundel ions in bulk water. At the very early stage of the hydration encompassed by our simulation, silanol groups are very unstable and molecular adsorption of water is slightly more stable than dissociative adsorption.
\end{abstract}

\begin{keyword} Tricalcium silicate. Hydration. \emph{ab initio} Molecular Dynamics. Proton
transfer. Interface. \end{keyword}

\end{frontmatter}


\section{Introduction} 

Although the use of low clinker ratio cements is increasing, the development of
new clinker types remains a reliable strategy to reduce greenhouse gas emissions
and improve the properties of cement for concrete structures applications. In a
context of durable design, considerable efforts are being made for a better
understanding of original Portland cement (OPC) hydration. However, considering
the hydration of all clinker phases together would turn the study highly
complex. Most of the time, the tricalcium silicate (\ce{C3S}) received a
particular attention, due to its predominance in OPC clinker (about
\SI{50}{\percent} to \SI{70}{\percent} by mass). \ce{C3S} is the main phase
responsible for OPC setting and strength development. The reactivity and early
hydration of tricalcium silicate (\ce{C3S}) is a very relevant topic towards a more sustainable design of OPC. The hydration itself
encompasses several processes such as dissolution, phase growth, diffusion and
complexation \cite{Bullard2011}. Atomistic simulation methods have shown good
capabilities in predicting the reactivity of mineral surfaces and the behavior
of solid/liquid interfaces. Towards a better understanding of the phases and processes occurring in cementitious systems, many atomistic force fields have been optimized \cite{Mishra2017}.  Molecular dynamics (MD) and density
functional theory (DFT), have already been used to compute surface energies and
determine Wulff shapes for monoclinic \ce{M3 C3S}
\cite{Manzano2015,Durgun2014,Mishra2013}. In recent studies, molecular and
dissociative adsorption of single water molecule were investigated on multiple
surface planes of \ce{M3 C3S} polymorph \cite{Zhang2018b,Zhang2019,Qi2019}.
Zhang et al. have shown that the adsorption energy decreases with increasing
amount of adsorbed molecules. Reactive MD studies indicates that after
approximately \SI{0.3}{\ns}, the structural properties of the surface are lost,
making further hydration process independent of the crystallographic surface
plane, and driven by proton hopping mechanisms towards the bulk
\cite{Manzano2015,Huang2015}. No correlation was found between water adsorption
energy and surface energy, when using static computational methods
\cite{Manzano2015}. However, the proton diffusion after the initial stage of
hydration was related to the location of the valence band maximum (VBM), which
is mainly constituted of oxygen 2p orbitals \cite{Huang2015,Huang2014}. Previous
DFT studies reveal that the local density of state of the VBM is close to the
oxygen anions for \ce{C3S}, and close to oxygen in silicates for \ce{C2S}
\cite{Durgun2012}. The higher reactivity of \ce{C3S}
when compared to \ce{C2S} is explained by the difference in their electronic
structure, arising from the presence of oxygen anions in \ce{C3S}. Calculations
of a single water molecule sorption on a (100) surface of \ce{T1 C3S}, shown
that chemisorption occured only in regions close to oxide ions. This behaviour
was associated to the higher degree of freedom of oxide ions when compared to
oxygen in silicate \cite{Huang2015}.

Proton transfer (PT) frequency strongly depends on hydrogen bonds (HB)
fluctuation due to thermal motion \cite{Tocci2014}, and thus cannot be analyzed
by a \SI{0}{\kelvin}, DFT investigation. Furthermore, such a phenomenon cannot
be captured considering a single water molecule adsorption. A previous
computational study found structural changes, as well as a huge increase in PT
rate from a solid/water monolayer interface to a thicker water film
\cite{Tocci2014}. Towards a better understanding of the \ce{C3S}/water
interface, we performed an \emph{ab initio} MD (AIMD) simulation, considering a
water film thick enough to account for fluctuation of the HB network. AIMD is a
powerful tool that has been used extensively to investigate the structural and
dynamical behavior of water/oxide interfaces at the DFT level of theory
\cite{Bjoerneholm2016,Tocci2014,Ngouana-Wakou2017,Rimsza2016,Cimas2014,Laporte2015,Lee2015,Liu2008}.
However, only few AIMD studies were conducted on cementitious materials
\cite{Churakov2009,Churakov2009a,Churakov2017}. As far as we know, this is the
first time that the very early hydration stage of \ce{C3S} is investigated using
AIMD. In particular, the structure of water and the PT dynamics are analysed and
quantified, and the results are compared with reactive molecular dynamics
calculations, performed for that purpose.

\section{Computational Methods}

\begin{figure}
\centering
\includegraphics{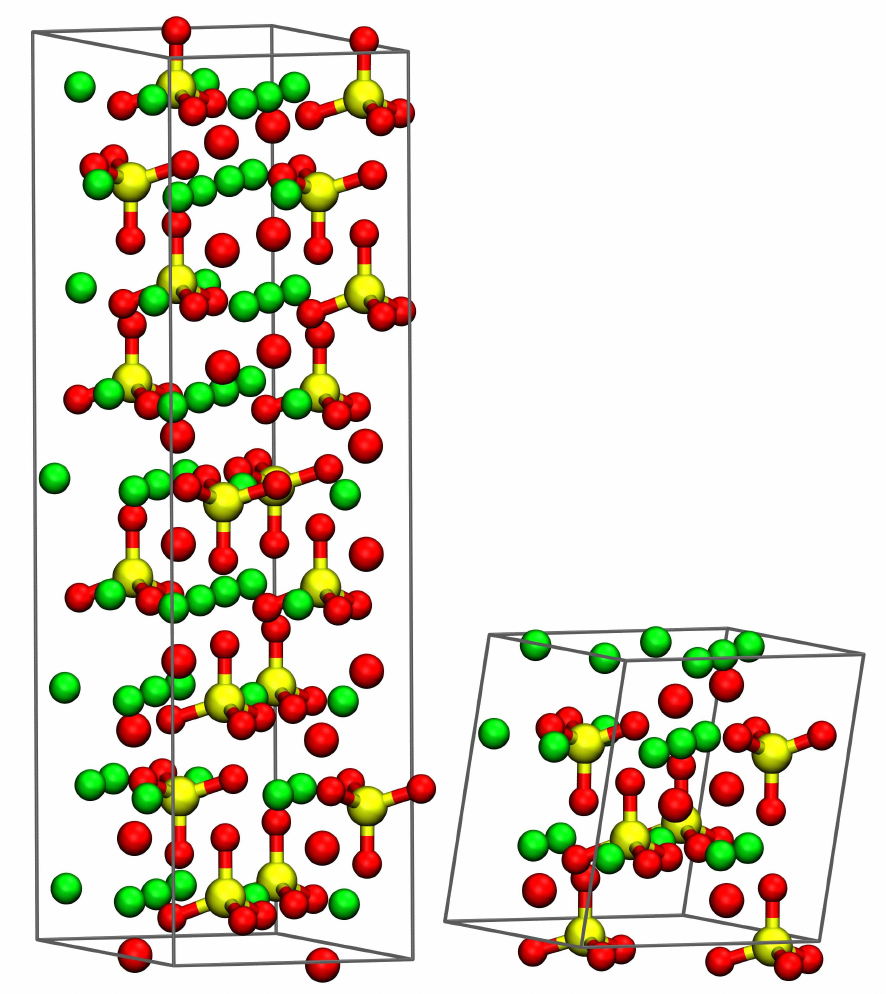}
\caption{\ce{M3} monoclinic cell, optimized from the model by Mumme et al.
\cite{Mumme1995} (right). Corresponding orthorhombic supercell (left). Color
code: calcium cations in green, oxygen  anions and silicate oxygen in red,
silicon atoms in yellow.}
\label{fig:cells}
\end{figure}

A simulation of the \ce{C3S}/water interface was performed on the symmetric,
Ca-rich, (040) plane (as in \cite{Mishra2013}). The \ce{M3 C3S} model employed
was refined from XRD analysis by Mumme et al. \cite{Mumme1995}. The unit
cell of 54 atoms was optimized, at the DFT level with the Quantum Espresso code,
using the PBE exchange-correlation functional \cite{Perdew1996,Perdew1997} with
a Grimme D2 correction for van der Waals interactions \cite{Grimme2006}. The
kinetic energy cutoffs for wave functions and charge density were
\SI{45}{\rydberg} and \SI{405}{\rydberg}, respectively. The Monkhorst-Pack
method was used for the integration of the first Brillouin zone, with a
\num{3x3x3} k-point mesh. During the optimization process, the atoms were
allowed to relax. In order to build a surface model, the optimized unit cell was
converted to an orthorhombic supercell of 162 atoms, with lattice parameters $a$
= \SI{12.28}{\angstrom}, $b$ = \SI{7.09}{\angstrom} and $c$ =
\SI{25.59}{\angstrom}. This transformation was performed with the Atomsk code,
which searches linear combination of the unit cell vectors producing vectors
aligned with Cartesian axes \cite{Hirel2015}. The optimized monoclinic cell and
corresponding orthorhombic supercell are represented in \cref{fig:cells}.

\begin{figure}
\centering
\includegraphics{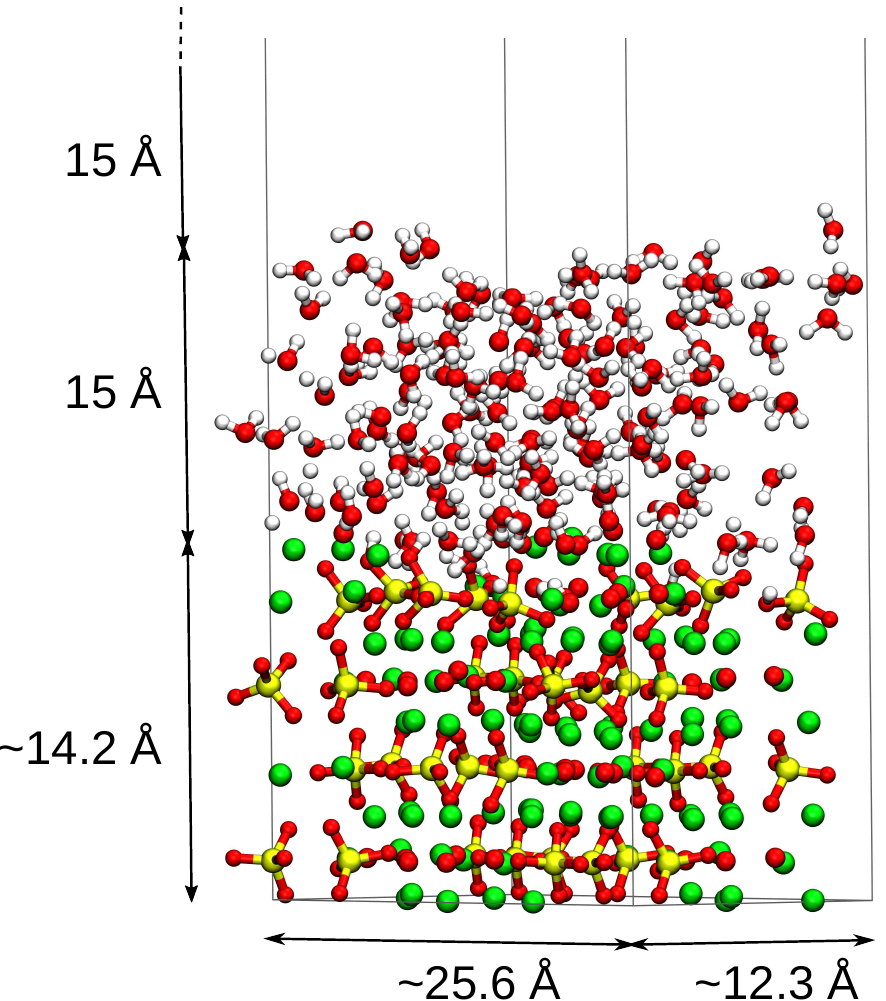}
\caption{Structure of the investigated \ce{C3S}(040)/water interface.}
\label{fig:system}
\end{figure}

The surface model was created from three orthorhombic supercells, with a
\SI{20}{\angstrom} thick vacuum layer, thus resulting in a \SI{12.28 x 25.59
x 21.28}{\angstrom} structure. The relaxation of the surface and the AIMD
simulation were performed with the CP2K code, using a PBE functional, and a
combination of Gaussian and plane wave basis functions (GPW), with Grimme D2
correction. A \SI{400}{\rydberg} planewave cutoff was adopted, and the
reciprocal space was sampled only at the $\gamma$ point. To relax the surface, the periodicity was applied for in-plane directions, and removed in the
direction of the vacuum. The atoms of the surface were allowed to relax at
the DFT level, ensuring that almost no change occurs within the middle of
the slab. The interface model was created adding a \SI{15}{\angstrom} thick
layer of water (157 molecules), with a \SI{15}{\angstrom} vacuum region. The
structure of the \ce{C3S}(040)/water interface is depited in \cref{fig:system}.

While the atoms of the mineral surface were kept fixed, the water molecules were
allowed to relax on the surface during a \SI{2}{ns} classical MD run in NVT
ensemble at \SI{300}{\kelvin}, using the INTERFACE FF parameters for \ce{C3S}
\cite{Mishra2013} and a SPC model for water \cite{Berendsen1981}. In order to
minimize the computational time, the bottom layer was removed so that the
remaining slab was composed of two orthorhombic supercells (\SI{\sim
14}{\angstrom} thick) and during the AIMD run, the lower supercell, considered
as the bulk, was fixed. Afterwards, a \SI{18}{\ps} AIMD run was performed within
the Born-Oppenheimer approximation, in the canonical ensemble, with a
Nose-Hoover thermostat, integrating the equation of motion with a \SI{0.5}{\fs}
timestep. Based on the evolution of the energy of the system, it was considered
that the equilibrium was reached after \SI{6}{\ps}, and the remaining
simulation time was used for analysis of equilibrium properties. A slightly higher temperature of
\SI{360}{\kelvin} (compared to standard conditions) was used to balance the low
diffusivity of water using the PBE functional. Deuterium masses were used for
protons to minimize the vibrational frequency of nuclei. It deserves to highlight that such substitution could decrease the frequency of PT, because of the lower vibrational frequency of deuterium nuclei in comparison to hydrogen nuclei. However, this method has already been used in the literature to prevent from energy drifts and it is understood that its benefits outweigh losses
\cite{Leung2006,Tocci2014}. A reactive molecular dynamics simulation was
performed using the ReaxFF, with the current optimized set of parameters for
Ca/O/H/Si elements \cite{Fogarty2010,Manzano2012,Manzano2012a}. The simulation
method and parameters, as well as the system size, are in accordance with
previously reported calculations \cite{Manzano2015}.

\section{Results}
\subsection{Water structure}

\begin{figure}
	\centering
	\includegraphics[width=\columnwidth]{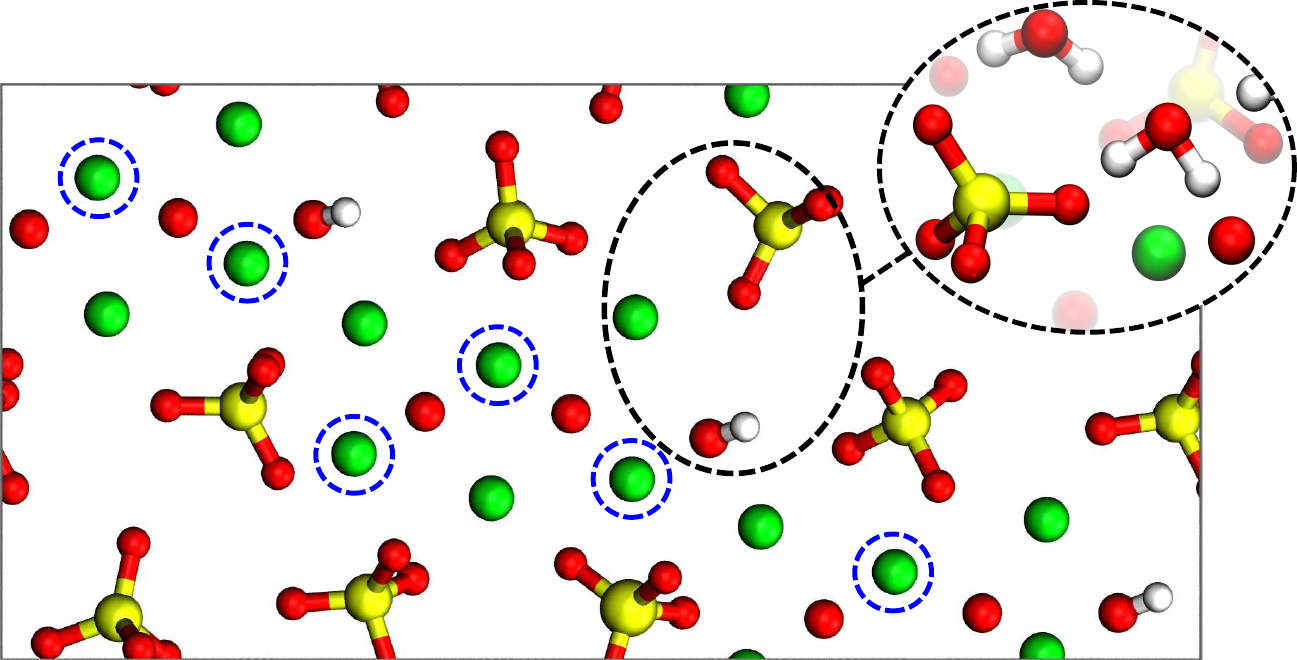}
	\caption{Atomic structure of the (040) surface. Hydrogen in hydroxide H-Oi are colorized in white. The snapshot shows the position of the water
	molecule before protonation of the oxide ion.  The ions of the surface are
	in the same layer, except for the calcium indicated by blue circles, having
	perpendicular coordinate $z$ \SI{\sim 2}{\angstrom} larger.}
	\label{fig:oxide_protonation}
\end{figure}

In this article, Oi refers to oxygen anions, Os refers to oxygens in silicates,
and Odw refers to oxygens resulting from the dissociation of water molecules.
At the very first steps of the simulation, three oxide ions Oi from equivalent
sites are protonated. The hydration model for \ce{C3S} proposed by
Pustovgar et al. considers that protonation of oxide ions occurs before
protonation of silicates \cite{Pustovgar2017}. Although hydroxides are more
stable than silanol groups, our simulation indicates that on the considered
(040) Ca-rich surface, protonation of Oi only occurs on sites close to
silicates. The other superficial oxide ions are shielded by four
calcium cations, hindering any protonation reaction. The negatively charged region allows water molecules to form hydrogen bonds with oxide ions on one side and with oxygens in silicates on the other side, thus leading to protonation of Oi (see
\cref{fig:oxide_protonation}).

\begin{figure}[H]
	\centering
	\includegraphics{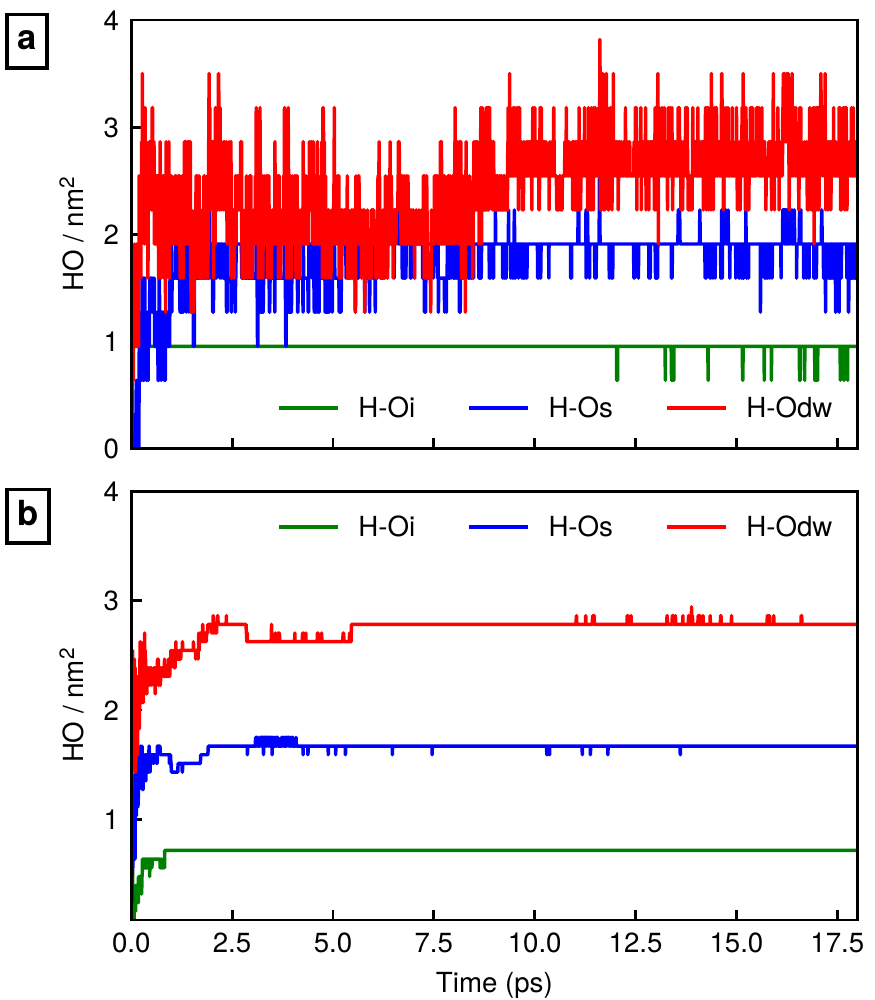}
	\caption{Number of hydroxyl groups formed at the surface on oxide ions
	(Oi-H), oxygen in silicates (Os-H), and from water dissociation (Odw-H) for
	(a) AIMD simulation, (b) ReaxFF simulation.}
	\label{fig:ho-coverage}
\end{figure}

The number of H-Os groups formed on silicates tends to stabilize after \SI{\sim
1}{\ps} whereas the number of H-Odw groups stabilize after \SI{\sim 0.25}{\ps}
(see \cref{fig:ho-coverage}). Both hydroxyl groups fluctuate during the whole
simulation due to proton transfer. Conversely, the hydroxides H-Oi formed on
oxygen anions Oi are very stable and no backward PT occurs. From our simulation
using the ReaxFF, within the timescale of \SI{18}{\ps}, a steady state is
reached very quickly as in the AIMD simulation. However, all hydroxyl groups
formed in the ReaxFF simulation are very stable, and the currently developed set
of parameters for Ca/O/H/Si failed to describe the PT dynamics between water and
Oi/Os atoms which is observed in our AIMD simulation. The complexity to optimize
parameters for ReaxFF relies on the fact that the same set of parameters is
employed for each element \cite{Senftle2016}. It means that the parameters are
the same independently of the environment. Other approaches based on the
empirical valence bond model has succeeded in reproducing \ce{OH-} solvation and
transport in water solutions \cite{Ufimtsev2009}. The implementation of a PT
model almost doubled the diffusion of \ce{OH-} ions when compared to a classical
model. The hydroxyl coverage of each hydroxyl type is close to the result of the
AIMD simulation. Therefore, within the timescale of the simulation, the ReaxFF
is representing the protonation state of the (040) surface in good agreement
with the AIMD simulation. The average total hydroxyl coverage over the last
\SI{12}{\ps} of simulation is \SI{5.36(37)}{\hopernmsquared} for the AIMD
simulation, and \SI{5.17(1)}{\hopernmsquared} for the ReaxFF simulation. These
values are also in agreement with previous investigation on \ce{C3S} hydration,
using the ReaxFF \cite{Manzano2015,Huang2015}.

\begin{figure*}
	\centering
	\includegraphics{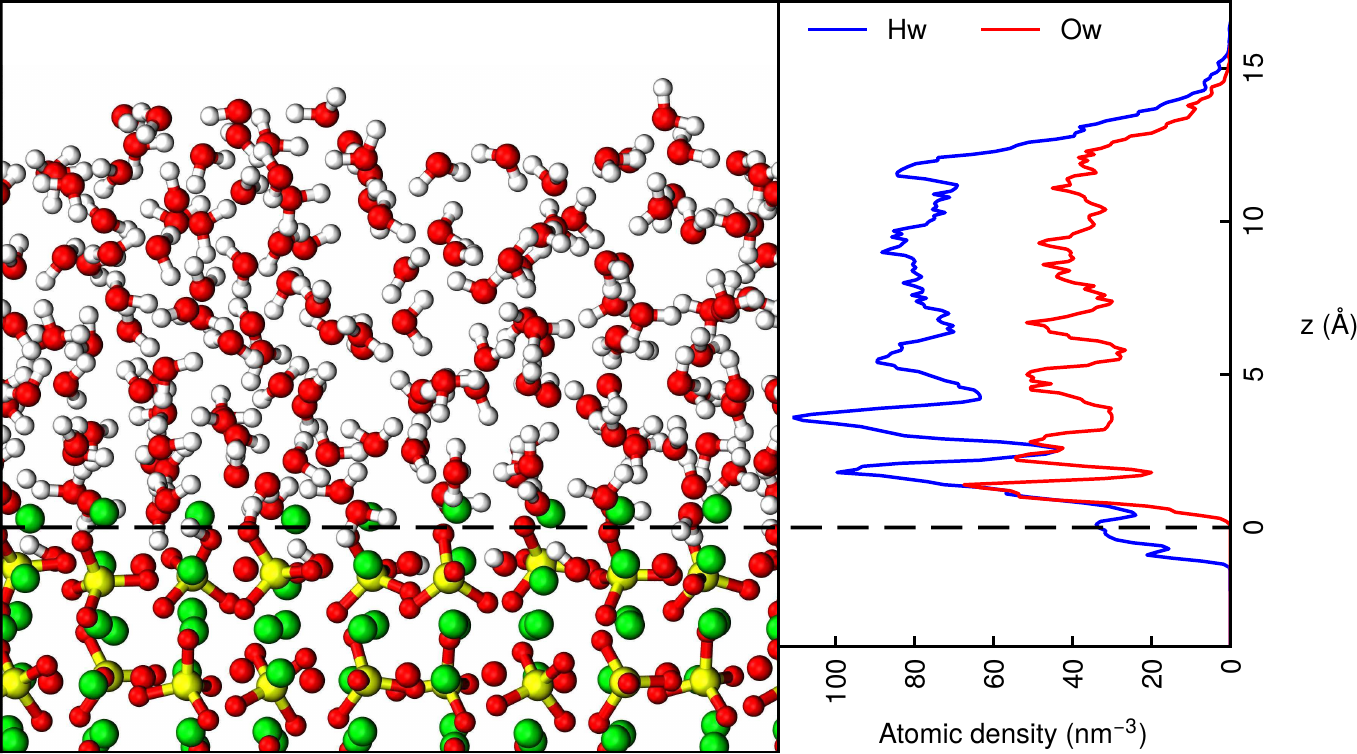}
	\caption{Atomic density profile of water molecules along the z axis for
	AIMD. The origin adopted is the average coordinate of the uppermost oxygen
	silicate layer. The layering observed results from the hydrogen bond network
	created between the strong interaction between the water molecules and the
	ionic surface.}
	\label{fig:1d_density_snap}
\end{figure*}

The atomic density profile of water oxygen and hydrogen atom, along the axis
perpendicular to the surface, is reported in \cref{fig:1d_density_snap}.
The layered structure of the interfacial water results from the effect of
excluded volume, electrostatic force field, and hydrogen bonding network.
Previous investigations based on classical MD simulations showed that this
layering is lost with protonation of the surface \cite{Claverie2019}. The
closest hydrogen's peak from the surface, is at the average position of oxygen
in silicates Os ($z = 0$), and corresponds to chemisorbed H. The thickness of
the layered region is approximately the same as in our previous classical MD
investigation: \SIrange{\sim 5}{6}{\angstrom} \cite{Claverie2019}. 

\begin{figure}[H]
	\centering
	\includegraphics{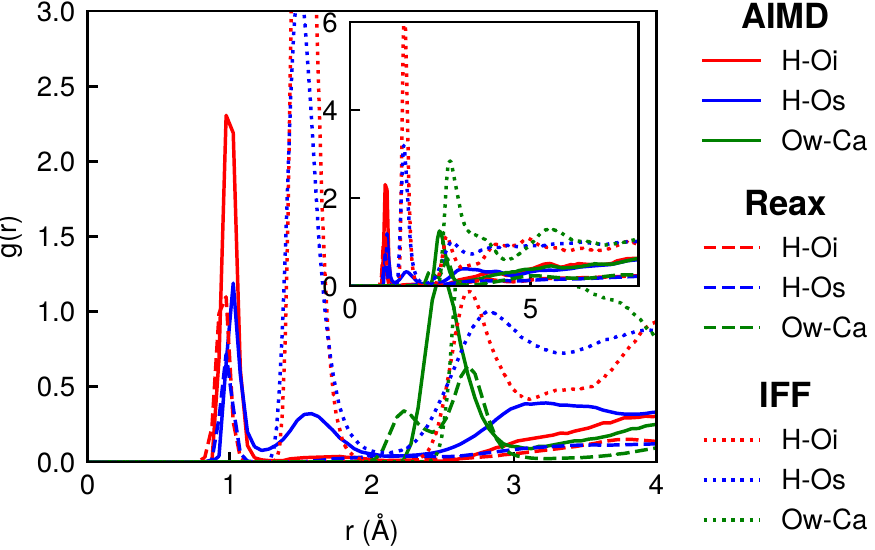}
	\caption{Radial distribution function H-Oi, H-Os, and Ow-Ca pairs employing AIMD, ReaxFF, and IFF \cite{Claverie2019}. The inset shows a zoom out ($\times 0.5$).}
	\label{fig:rdf}
\end{figure}

The radial distribution functions (RDF) of H-Oi, H-Os and Ow-Ca pairs are
plotted in \cref{fig:rdf}. 
For H-Oi and Ow-Ca pairs, sharp peaks raise at \SI{\sim 0.97}{\angstrom} and
\SI{\sim 2.50}{\angstrom} respectively, indicating superficial hydroxides and
the first coordination shell of calcium cation. No second coordination shell is
noticed for both pairs. Yet for the H-Os pairs, two peaks stand at \SI{\sim
1.02}{\angstrom} and \SI{\sim 1.58}{\angstrom}, corresponding to hydroxyl groups
formed on silicates and H-bonds between water and oxygen in silicates. The RDF
obtained in the ReaxFF simulation is very similar, with two main differences: in
one case, only one correlation peak is observed for the H- Os pair, suggesting
an Os protonation, and the coordination peak for Ow-Ca pairs is split in two,
indicating two different distances of correlated water molecules. RDF for dry
\ce{C3S}/water interface, obtained from previous MD investigation
\cite{Claverie2019}, was plotted for comparative purpose. In such classical
simulation, no PT occurs and the first coordination peak correspond to H-bonds
between water and superficial anions (r \SI{\sim 1.53}{\angstrom}).

\begin{figure}[H]
	\centering
	\includegraphics[width=\columnwidth]{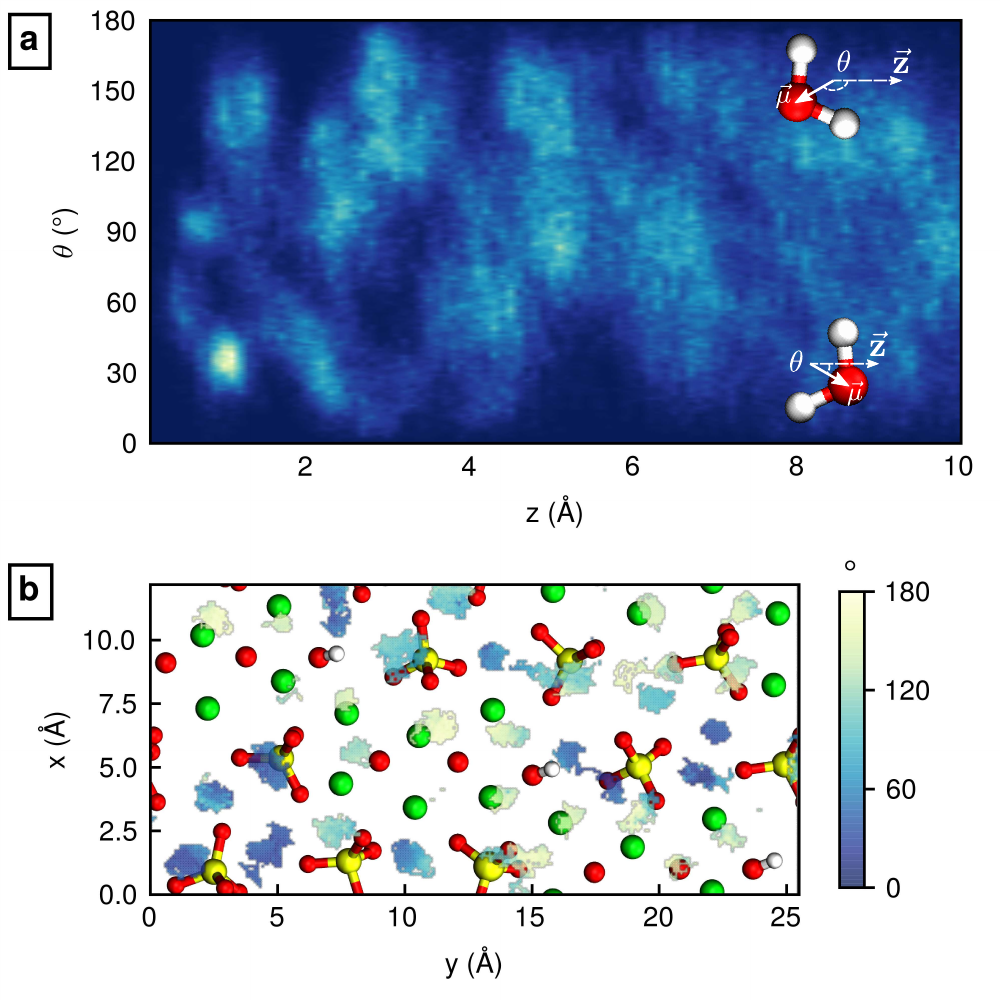}
	\caption{(a) Probability distribution of the angle $\theta$ between the
	water dipole moment and the z axis, perpendicular to the surface. Lighter
	regions correspond to higher probabilities (b) Density mapping of $\theta$
	for water molecules within \SI{3}{\angstrom} from the surface. Color code:
	Ca in green, Os and Oi in red, Si in yellow, H in hydroxide H-Oi in
	white.}
	\label{fig:angdis_water}
\end{figure}

\begin{figure}[H]
\centering
\includegraphics[width=\columnwidth]{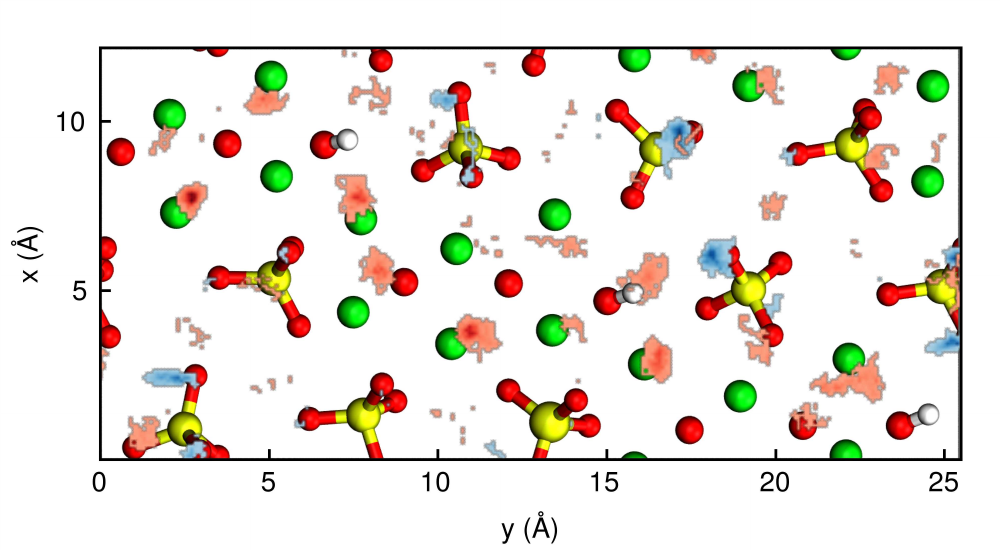}
\caption{Normalized probability distribution of H-Os (blue) and H-Odw (red) hydroxyl groups.}
\label{fig:map_oh}
\end{figure}

The orientation of water molecules in contact with the surface is a
characteristic of the hydrophilic/hydrophobic behavior of the surface. The
probability distribution of the angle $\theta$ between the water dipole moment
and the $z$ axis is depicted in \cref{fig:angdis_water} a). Within the contact
layer ($z$ \SI{< 1.6}{\angstrom}), most of the water molecules have $\theta$
\SIrange[range-phrase = --]{\sim 20}{50}{\degree} or $\theta$
\SIrange[range-phrase = --]{\sim 120}{160}{\degree}, meaning that their dipole
moments point preferentially towards or against the surface. This feature is
characteristic of hydrophilic surfaces \cite{Deryagin1987}. Less probably, water
molecules forming a single hydrogen bond with silicates orient with their dipole
moment parallel to the surface.   Water molecules in the contact layer orient
according to the charge of superficial ionic species. Thus, two regions can be
distinguished: one where the water dipole is oriented upward and H atoms
coordinate with Os, and a second where the water dipole is oriented downward and
Ow atoms coordinate with calcium cations. These regions are mapped on the
surface in \cref{fig:angdis_water} b) by collecting $\theta$ and the $x$ and $y$
coordinates of water molecules within \SI{3}{\angstrom} from the surface, during
the whole simulation.  The effect of the topology of \ce{C3S} surfaces on the
structure of water molecules has already been reported in a previous molecular
dynamics investigation \cite{Alex2017}.
The probability distribution of H-Os and H-Odw hydroxyl groups is plotted in
\cref{fig:map_oh}. H-Odw groups are principally located on Ca-rich, positively
charged regions, but also close to protonated silicates.

\subsection{Proton transfer analysis}

\begin{figure}[H]
	\centering
	\includegraphics[width=\columnwidth]{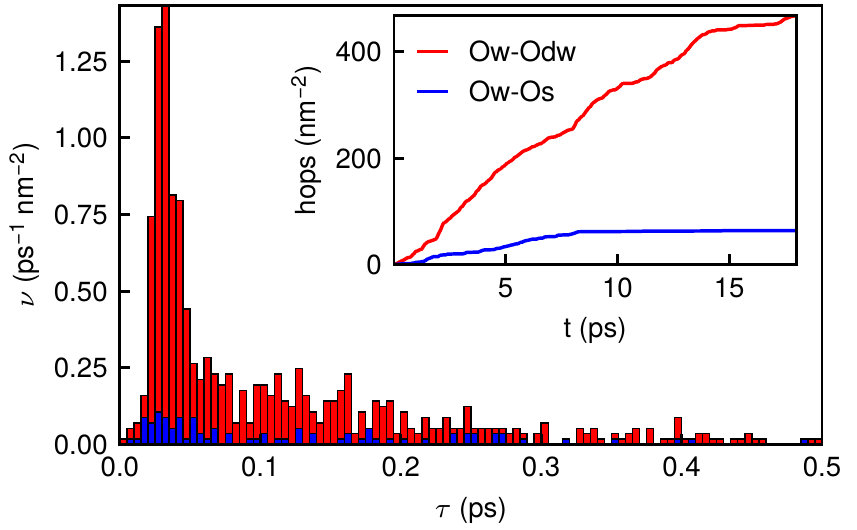}
	\caption{Frequency $\nu$ of PT as a function of its lifetime $\tau$ (left). Evolution of the total number of hops within the simulation time (right).}
	\label{fig:frequency}
\end{figure}

The frequency $\nu$ of PT between water molecules and Os-H and Odw-H groups is
reported in \cref{fig:frequency}. The lifetime $\tau$ is defined as
the time for a proton to return to the oxygen atoms to which it was initially
bonded.
The inset plot shows the evolution of proton hops during the simulation. Hops
between Ow and Os reach a plateau after \SI{\sim 8}{\ps}. Conversely, hops
within Odw-Ow pairs seems to increase steadily during the whole simulation and
are about 5 times more frequent than within Os-Ow pairs during the first
\SI{8}{\ps}. The typical lifetime of PT events is shorter than
\SI{100}{\fs}, which corresponds to the time scale of Eigen-Zendel structure
interchanging in bulk water, obtained by femtosecond vibrational spectroscopy
\cite{Woutersen2006}. Lifetimes of approximately the same duration were reported
for PT at water/ZnO($10\bar{1}0$) interface \cite{Tocci2014}. The longest
lifetimes reported are \SI{\sim 0.2}{\ps} for transfer between Ow and Os, and
\SI{\sim 2}{\ps} for transfer between Ow and Odw.

\begin{figure*}
	\centering
	\includegraphics{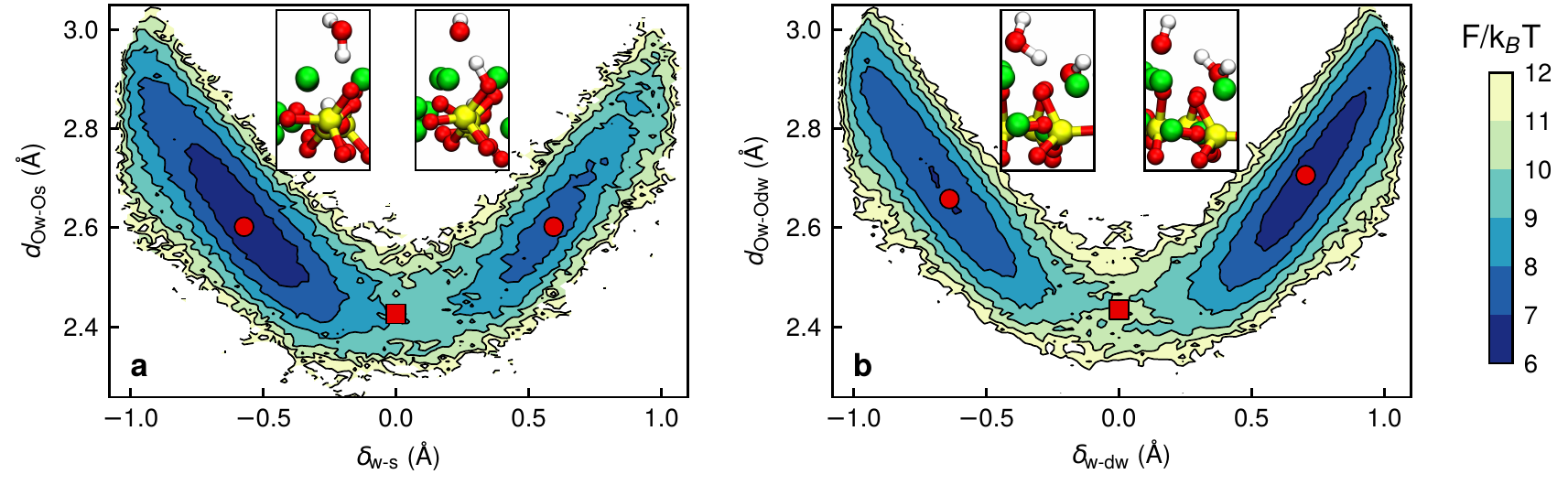}
	\caption{Free energy contour plot of PT between Ow and Os as a function of
	the distance $d_{\text{Oa-Ob}}$ between oxygen atoms, and of
	$\delta_{\text{a-b}} = d_{\text{Oa-H}} - d_{\text{Ob-H}}$. The red circles
	shows the local free energy minima, corresponding to the most stable
	configurations before and after the proton jump. The red square is the
	saddle point, where the proton is equidistant from both oxygen atoms.}
	\label{fig:free-energy}
\end{figure*}

The free energy profiles of PT between Ow and Os and between Ow and Odw pairs
were obtained from a standard method as follows: $F = -k_BT \log P$ (see \cref{fig:free-energy}) \cite{Tocci2014, Ufimtsev2009, Zhu2002}.
$P$ is a probability distribution function of the distance $d_{\text{Oa-Ob}}$
between two oxygen atoms Oa and Ob, and of $\delta_{\text{a-b}} =
d_{\text{Oa-H}} - d_{\text{Ob-H}}$. In the case of PT between Ow and Os, the
molecular configuration of water forming an H-bond with Os is more stable than
the dissociative configuration. The free energy profile of PT between Ow and Odw
suggests that the molecular adsorption of water on the surface is more stable
than the dissociated form. This observation is contrary to previous DFT and
reactive MD calculations on a single water molecule, where dissociative
adsorption energies were generally larger than molecular adsorption energies,
resulting in more stable configurations \cite{Zhang2019, Zhang2018b,Qi2019,
Manzano2015}. This consolidates the idea that static calculations on single
molecule adsorption cannot describe accurately the properties of a solid/water
interface. Tocci and Michaelides also reported considerable differences in PT
rate between a monolayer and an thicker water film \cite{Tocci2014}. These
differences arise from the decrease of the free energy barrier of proton
transfers induced by H-bond fluctuations. The free energy barrier is \SI{\sim
3.24}{\kt} for PT from Ow to Os and \SI{\sim 2.24}{\kt} from Os to Ow. The
barrier is \SI{\sim 3.30}{\kt} for transfer from Ow to Odw and \SI{\sim
3.60}{\kt} for the reverse reaction. These results suggest that hydroxides
formed by water dissociation are more stable than silanol groups. However, the
energy barrier for the creation of both hydroxyl groups is almost equal.

Electron density difference analysis allows to map the distribution of electrons
involved in PT, and more generally in the adsorption. This analysis was realized
performing static DFT calculation on the total system, as well as \ce{C3S}, and
water independently on the system configuration at \SI{5}{\ps}. The electron
density difference $\Delta \rho$ was calculated as follow:
\begin{equation}
\Delta \rho = \rho_{\ce{C3S}/\ce{H2O}} - \rho_{\ce{C3S}} -\rho_{\ce{H2O}}
\end{equation}
where $\rho_{\ce{C3S}/\ce{H2O}}$ corresponds to the electron density of the
interface system, $\rho_{\ce{C3S}}$ and $\rho_{\ce{H2O}}$ are the electron
density of the \ce{C3S} and water alone, respectively. A positive value of
$\Delta \rho$ indicates a high electron density, while a
negative value points out an electron depletion region.

\begin{figure*}
	\centering
	\includegraphics{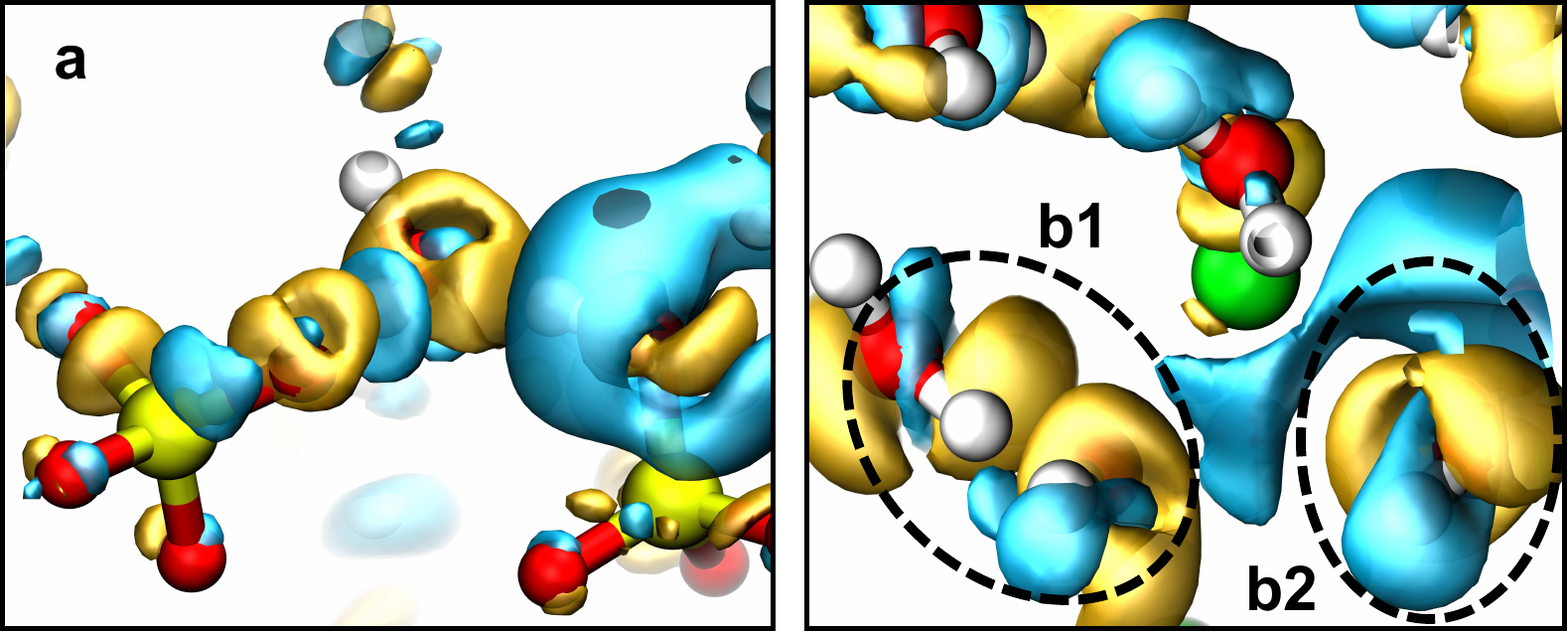}
	\caption{Snapshots of the isosurface of electron density difference. Golden and cyan isosurface represent positive and negative $\Delta \rho$, respectively. a) A water molecule between a silicate and a silanol group.  b1) Electron density delocalization during a proton exchange between a molecular water and a hydroxide. b2) Electron abundance and depletion regions around an hydroxide created upon water dissociation.}
	\label{fig:edensity}
\end{figure*}

The proton transfers occurring at the surface create an electron
delocalization, and thus high and low electron density regions. Water molecules
act as charge carrier and the electron depletion or gain highly depends on the
location of the molecule (see \cref{fig:edensity} a and b1). High electron
density is observed around the silicate oxygen close to the water molecule,
creating a depletion region on the silicon atom. A large depletion region is
observed around the silanol group (\cref{fig:edensity} a). Charge depletion
regions are observed around H of hydroxyl groups. Their magnitude increase in
this order: H-Oi $<$ H-Odw $<$ H-Os. In other words, the magnitude of the
depletion regions on H decreases as a function of the stability of the hydroxyl
group. A large depletion region indicates a greater charge separation, and a
more ionic bond, while a small depletion region reveals a more covalent bond.

\section{Conclusion}
The very early hydration of the (040) surface of \ce{C3S} was investigated
through a \SI{18}{\ps} AIMD simulation. As a first observation, only 1/3 of the
oxide ions on the surface were protonated during the whole simulation. The
hydroxides formed are highly stable and no proton exchange was observed.
Although the oxide ion is very unstable in water, we found that its environment
on the surface is an important factor for the creation of hydrogen bond with
water molecules and for protonation to occur. Thus, the pKa of hydroxide and
silicic acid in solution cannot predict accurately the protonation state of the
surface during the very early stage of hydration. The structure of water at
the interface, resulting from the formation of the hydrogen bond network, is
very similar to that of our previous classical molecular dynamics study, with a
thickness of the layered region of approximately \SIrange{5}{6}{\angstrom} from
the surface. The (040) surface is composed of Ca-rich regions (positively
charged) and Si-rich regions (negatively charged). Water molecules in the
contact layer orient their dipole moment in accordance with the surface charge,
making either H-bonds in Si-rich regions, or creating strong Ca-Ow interactions
in Ca-rich regions. Energy barrier analysis suggests that the molecular
adsorption of water on the \ce{C3S} surface is more stable than dissociative
adsorption. Based on proton transfer energy analysis, the hydroxyl groups formed
were classified in order of stability as follow: H-Oi $>$ H-Odw $>$ H-Os. From
electron density difference, high electron density, and depletion regions were
observed. These observations revealed that the magnitude of the electron
depletion region upon adsorption is smaller for more stable hydroxyl groups. 

\section*{Acknowledgments}
The authors acknowledge Brazilian science agencies CAPES (PDSE process
n\si{\degree}88881.188619/2018-01) and CNPq for financial support, as well as
Hegoi Manzano and Gabriele Tocci for helpful discussions. The authors also thank
the anonymous reviewers for their careful reading of our manuscript and their
many insightful comments and suggestions.
	
\bibliographystyle{elsarticle-num}
\biboptions{sort&compress}
\bibliography{ccr_2019}

\end{document}